\documentclass[12pt]{article}
\usepackage{amsmath,amssymb,epsfig}

\def\be{\begin{eqnarray}}
\def\ee{\end{eqnarray}}

\begin{document}

\begin{center}
{\Large \bf Nonforward anomalous dimensions  of Wilson operators  \\
in $N=4$ super-Yang-Mills theory. }
\\ \vspace*{5mm} A.I.~Onishchenko$^{a,b}$
and V.N.~Velizhanin$^{c}$
\end{center}

\begin{center}
a) Department of Physics and Astronomy \\ Wayne State University,
Detroit, MI 48201, USA \\
\vspace*{0.5cm}
b) Institute for Theoretical and Experimental
Physics, \\ Moscow, Russia \\
\vspace*{0.5cm}
c) Theoretical Physics Department, \\ Petersburg Nuclear Physics Institute \\
Orlova Rosha, Gatchina, \\
188300, St. Petersburg, Russia
\end{center}

\begin{abstract}
We present the next-to-leading order results for universal non-forward anomalous dimensions of
Wilson twist-2 operators in N=4 supersymmetric Yang-Mills theory. The
whole calculation was performed using supersymmetric Ward identities
derived in this paper together with already known QCD results and does
not involve any additional calculation of diagrams. We also considered
one particular limit of our result, which could potentially be interesting
in the context of AdS/CFT correspondence.
\end{abstract}

\section{Introduction}

Parton distributions in QCD satisfy the
Balitsky-Fadin-Kuraev-Lipatov (BFKL)~\cite{BFKL} and
Dokshitzer-Gribov-Lipatov-Altarelli-Parizi (DGLAP)~\cite{DGLAP} equations.
Up to day even next-to-leading corrections to these equations are
known~\cite{NLOBFKL,NLOAD}. Moreover, recently, similar results were also
obtained for supersymmetric theories~\cite{KL00,KL,KLV}. An idea to
consider these equations in the case of supersymmetric theories is
based on common expectations, that the presence of extra high
symmetry may significantly simplify them as well as their
analysis. For example, it was already known, that all
quasi-partonic operators in $N=1$ SYM form supermultiplet of
operators~\cite{BFKL2}, having the same single universal anomalous dimension
with shifted argument, which at that time was computed in leading order (LO) of
perturbation theory. Calculations in $N=4$ SYM gave even more
remarkable results -
the eigenvalues of the integral kernels in the
evolution equations for quasi-partonic operators
are proportional to $\Psi (j-1)-\Psi(1)$~\cite{N=4,LN4}, which means that these
evolution equations in the multicolour limit are equivalent to the
Schr\"{o}dinger equation for the integrable Heisenberg spin model~\cite{N=4}
similar to the one found in the Regge limit~\cite{Integr}.
Moreover, it was shown, that in the maximally
supersymmetric $N=4$ Yang-Mills theory there is a deep relation
between BFKL and DGLAP evolution equations~\cite{KL}. In
particular, the anomalous dimensions of Wilson twist-2 operators
in $N=4$ SYM could be found from the eigenvalues of the kernel of
BFKL equation. The corresponding next-to-leading order (NLO)
calculations showed, that many of these findings are valid also in
higher orders of perturbation theory. Moreover, as in leading
order they most fully realize in maximally supersymmetric $N=4$
SYM.

However, some of the properties of these equations, valid at
leading order, are violated at higher orders. For example, the
conformal invariance of the theory in leading order allows us to
construct multiplicatively  renormalized quasi-partonic
operators~\cite{BFKL2} up to this order in perturbation theory.
But, in next-to-leading order in perturbation theory
multiplicative renormalization of conformal operators is violated
due to necessity in regularization of arising ultraviolet
divergences, which is responsible for violation of conformal
symmetry.

The Efremov-Radyushkin-Brodsky-Lepage (ER-BL) equation~\cite{ERBL}
may be viewed as some kind of the generalization of the DGLAP
equation for the case of non-forward distribution functions. In
this case in and out hadronic states in matrix elements of
operators are different, what allows us to study scale properties
of hadron wave functions. In the latter case the matrix elements
of corresponding operators are considered between vacuum and
hadronic state. In leading order the kernel of ER-BL evolution
equation, after Mellin transformation was performed, is diagonal
in basis of Gegenbauer polynomials~\cite{MRQCD,MRPhi36}. This is
equivalent to the fact, that multiplicatively renormalized
operators of ER-BL evolution kernel coincide in leading order with
conformal operators constructed from Gegenbauer polynomials~\cite{CPO}.
The direct calculation of next-to-leading order corrections to the evolution kernel of this
equation turned out to be much more difficult problem compared to
the forward case. It was first performed for the case of non-singlet quark
operator in~\cite{DR,MRQCD}, where in the last reference an advanced computation
method was developed  and some analysis of general
properties of evolution kernel and corresponding anomalous
dimension matrix was given. In this case as well as in the case of
$\phi^3_{[6]}$ model a solution of ER-BL evolution equation was
constructed at next-to-leading order of perturbation
theory~\cite{MRKernel}. It was noticed, that at next-to-leading order
conformal invariance is violated and anomalous dimension matrix
develops non-diagonal part in the basis of conformal operators.

The source of conformal symmetry breaking was identified later
in~\cite{MPhi3, MQCDNS}. It was found, that non-diagonal part of
anomalous dimension matrix arises entirely due to the violation of
special conformal symmetry. Moreover, a framework based on the
analysis of broken conformal Ward identities was proposed, which
allowed to obtain for the first time next-to-leading order
corrections to the nonforward anomalous dimension matrix in
singlet case both in  QED~\cite{BMQED} and in QCD~\cite{BMQCD}.
The generalization of these results for the case of supersymmetric
theories involves a number of relations between elements of anomalous
dimensions matrix~\cite{SUSYCWI}, some of which were already known
from the analysis of forward limit (last paper in Ref.~\cite{DGLAP},\cite{BFKL2}).
Later, the use of these supersymmetry relations, which could be also easily written
for the corresponding evolution kernel matrix~\cite{BFKL2,SUSYCWI}, allowed to
determine for the first time the QCD ER-BL evolution kernels in the singlet case
both for odd~\cite{BMOdd} and even~\cite{BMeven} parity distribution amplitudes
(see also~\cite{BFM}).

In this paper we, using already known results from QCD, found
closed analytical expression for universal non-forward anomalous
dimension for maximally supersymmetric $N=4$ Yang-Mills
theory. Earlier, the authors of ~\cite{KL} used already known QCD
results~\cite{NLOAD} to obtain expression for NLO universal forward anomalous
dimension of Wilson twist-2 operators in $N=4$ SYM, which were later
confirmed by direct calculations in Ref.~\cite{KLV}.
A problem, we are solving here, may be formulated as follows: how knowing
QCD results one may obtain analogous results in supersymmetric theory
with minimum efforts.

It turns out, that owing to remarkable properties of $N=4$ SYM it is possible
to derive an expression for universal non-forward anomalous dimension
without any additional calculation beyond those already done for QCD.

Our result could be also interesting in the context of  AdS/CFT
correspondence \cite{MGKPW}. Namely, there are some calculations of the
anomalous dimensions of such operators in the limit of large $j$
(Lorentz spin) from both sides of the AdS/CFT correspondence
\cite{GKP,FT,K2002,M2002,KLV,BGK}.
There are, also some predictions for anomalous
dimensions of other types of operators corresponding to multi-spin
solutions in $AdS_5\times S_5$ space from string theory
side~\cite{FT03} \footnote{see last reference in~\cite{FT03} for review},
partially confirmed by field theory calculations.

It should be noted, that up to this moment only diagonal part of
anomalous dimension matrix have been studied in the context of AdS/CFT
correspondence and it would be interesting to compare the results for
its non-diagonal part with the appropriate result from string theory.

The present paper is organized as follows. First in next two subsections
we define the lagrangian of $N=4$ SYM and introduce singlet conformal
operators present in this theory. Next, in section 2 we derive supersymmetric
Ward identities and resulting constrains on anomalous dimensions of our
conformal operators. Then, in section 3 we proceed with the determination
of universal non-forward anomalous dimension for Wilson twist-2 operators
in $N=4$ SYM. And, finally, section 4 contains our conclusion.

\subsection{Lagrangian of $N=4$ SUSY YM}

Different supersymmetric Yang-Mills theories in four dimension
can be constructed from higher dimensional $N=1$ supersymmetric Yang-Mills
theory through dimensional reduction~\cite{BSS}. To obtain the lagrangian
of $N=4$ SYM in 4-dimensions we start with $N=1$ SYM in 10-dimensions
\begin{equation}
{\cal L}=
-\frac{1}{4}G_{\mu\nu}^aG^{a\;\mu\nu}+\frac{i}{2}\bar\lambda^a
\Gamma_\mu D^\mu\lambda^a.
\end{equation}
Here $\Gamma_\mu$ are 10-dimensional Dirac gamma-matrices, $G^a_{\mu\nu}$ is gauge field strength,
$D^\mu$ is covariant derivative and $\lambda^a$ is Majorana-Weyl spinor. The supersymmetry
transformations are
\begin{eqnarray}
\delta^Q {\cal A}^a_\mu&=& i \; \bar\xi \Gamma_\mu \lambda^a,\nonumber\\
\delta^Q \lambda^a &=& \Sigma_{\mu\nu} G^{a\;\mu\nu}\xi\,,\label{SUSYTrD10}
\end{eqnarray}
where $\Sigma_{\mu\nu}=\frac{1}{4}[\Gamma_\mu,\Gamma_\nu]$. After dimensional
reduction to four dimensions we get the following lagrangian of $N=4$ supersymmetric Yang-Mills theory~\cite{GSO}
\begin{eqnarray}
{\cal L}&=&-\frac{1}{4}G^a_{\mu\nu}G^{a\;\mu\nu}+\frac{i}{2}\bar\lambda^a
\gamma_\mu D^\mu\lambda^a +\frac{1}{2}\left(D_\mu A^a_r\right)^2
+\frac{1}{2}\left(D_\mu B^a_r\right)^2\nonumber\\
&-&\frac{g}{2}f^{abc}\bar\lambda^a
\left(\alpha_r A_r^b+\gamma_5 \beta_r B_r^b\right)\lambda^c\label{eq:1}\\
&-&\frac{g^2}{4}\left[\left(f^{abc}A_r^bA_t^c\right)^2
+\left(f^{abc}B_r^bB_t^c\right)^2
+2\left(f^{abc}A_r^bB_t^c\right)^2\right],\nonumber
\end{eqnarray}
where $\lambda^a$ denotes vector of 4 Majorana spinors and
\begin{eqnarray}
G_{\mu\nu}^a&=&\partial_\mu {\cal A}^a_\nu-\partial_\nu {\cal A}^a_\mu
+g f^{abc}{\cal A}_\mu^b {\cal A}_\nu^c,\nonumber\\
D_\mu\lambda^a&=&\partial_\mu\lambda^a+g f^{abc} {\cal A}_\mu^b\lambda^c,\nonumber
\end{eqnarray}
\begin{equation*}
\alpha_1=\left(
\begin{array}{cc}
 0 & \sigma_1   \\
 -\sigma_1 & 0
\end{array}\right),\qquad
\alpha_2=\left(
\begin{array}{cc}
 0 & -\sigma_3   \\
 \sigma_3 & 0
\end{array}\right),\qquad
\alpha_3=\left(
\begin{array}{cc}
 i\sigma_2 & 0  \\
 0 & i\sigma_2
\end{array}\right),
\end{equation*}
\begin{equation*}
\beta_1=\left(
\begin{array}{cc}
 0 & i\sigma_2   \\
 i\sigma_2 & 0
\end{array}\right),\qquad
\beta_2=\left(
\begin{array}{cc}
 0  & 1   \\
 -1 & 0
\end{array}\right),\qquad
\beta_3=\left(
\begin{array}{cc}
 -i\sigma_2 & 0  \\
 0 & i\sigma_2
\end{array}\right),
\end{equation*}
\begin{eqnarray}
\sigma_{\!\mu\nu}&=&\frac{1}{4}\left[\gamma_\mu,\gamma_\nu\right],\qquad
{\sigma '}_{\!\!rt}=\frac{1}{4}\left[\alpha_r,\alpha_t\right],\qquad
{\sigma}_{\!rt}=\frac{1}{4}\left[\beta_r,\beta_t\right],\qquad
{\kappa}_{rt}=\frac{1}{4}\left\{\alpha_r,\beta_t\right\}.\nonumber
\end{eqnarray}
In 4-dimensions we have gauge field (gluon), four Majorana fermions (gauginos),
three scalars ($A$) and three pseudoscalars ($B$). Our notation is somewhat
different form usually employed $SU(4)$-covariant form of $N=4$ SYM, where
we have gauge field, four left Weyl fermions and 6 real scalars. Certainly, both these
forms of $N=4$ lagrangian are completely equivalent and we have chosen this one only
because it seems to be more convenient in doing actual loop calculations.  The
4-dimensional supersymmetry transformation (\ref{SUSYTrD10}) now become
(here $\xi$ stands for a vector of 4 Majorana spinors)
\begin{eqnarray}
\delta {\cal A}_\mu^a&=&i\bar\xi\gamma_\mu\lambda^a,\nonumber\\
\delta A_r^a&=&\bar\xi\alpha_r\lambda^a,\nonumber\\
\delta B_r^a&=&\bar\xi\gamma_5\beta_r\lambda^a,\label{eq:2}\\
\delta \lambda^a&=&\left[\sigma_{\!\mu\nu}G^{a\;\mu\nu}\!\!+
i\gamma_\mu D^\mu\left(\alpha_r A_r^a+\gamma_5\beta_r B_r^a\right)\right.\nonumber\\
&&\hspace*{15mm}+\left.g f^{abc}\left({\sigma '}_{\!\!rt}A_r^b A_t^c+{\sigma}_{\!rt}B_r^b B_t^c
+2\gamma_5{\kappa}_{rt}A_r^b B_t^c\right)\right]\xi\,,\nonumber\\
\delta \bar\lambda^a&=&\bar\xi\left[-\sigma_{\!\mu\nu}G^{a\;\mu\nu}\!\!+
i\gamma_\mu D^\mu\left(\alpha_r A_r^a-\gamma_5\beta_r B_r^a\right)\right.\nonumber\\
&&\hspace*{21mm}- \left.g f^{abc}\left({\sigma '}_{\!\!rt}A_r^b A_t^c+{\sigma}_{\!rt}B_r^b B_t^c
-2\gamma_5{\kappa}_{rt}A_r^b B_t^c\right)\right]\,.\nonumber
\end{eqnarray}
Note, if we choose $\lambda_3=\lambda_4=0$, $\lambda_1$, $\lambda_2$
non-vanishing and $A_1=A_2=B_1=B_2=0$, $A_3=A$, $B_3=B$ non-vanishing,
then using the explicit representation of $\alpha$ and $\beta$ matrices
lagrangian (\ref{eq:1}) becomes lagrangian for $N=2$ SUSY theory.

Moreover, $N=4$ SYM has internal $SU(4)$ $R$-symmetry group, which is symmetry of
lagrangian under the following transformations of fields
\begin{eqnarray}
\delta {\mathcal A}_{\mu}^a &=& 0, \nonumber \\
\delta A_r^a &=& \Lambda^{'}_{rt}A_t^a + \widetilde{\Lambda}_{rt}B_t^a, \nonumber \\
\delta B_t^a &=& \Lambda_{rt}B_r^a - \widetilde{\Lambda}_{rt}A_r^a, \\
\delta\lambda^a &=& -\frac{1}{2}(
\sigma^{'}_{rt}\Lambda^{'}_{rt}
+ \sigma_{rt}\Lambda_{rt} + 2\gamma_5k_{rt}\widetilde{\Lambda}_{rt})\lambda^a, \nonumber \\
\delta\bar\lambda^a &=& \frac{1}{2}\bar\lambda^a(
\sigma^{'}_{rt}\Lambda^{'}_{rt} + \sigma_{rt}\Lambda_{rt} - 2\gamma_5k_{rt}\widetilde{\Lambda}_{rt}
), \nonumber
\end{eqnarray}
where  parameters of these transformations, given by real antisymmetric matrices $\Lambda^{'}_{rt}$,
$\Lambda_{rt}$ and $\widetilde{\Lambda}_{rt}$.

Later in this paper we will be interested in scale transformation properties
of Wilson operators of lowest twist or, what is the same, in those, which have maximal
Lorentz spin. The component with maximal Lorentz spin has a symmetric traceless
Lorentz structure and the simplest way to project it out is to use a convolution
with the product of light-like vectors $n_{\mu}$ $(n^2=0)$. This way we effectively project
our theory on the light-cone. For a general four-vector $X^\mu$ we introduce
light-cone coordinates as follows: $X^\pm=1/\sqrt{2}(X^0\pm X^3)$  and
$X^\mu Y_\mu=X^{+}Y^{-}+X^{-}Y^{+}-X^{i}Y^{i}$, where $i=1,2$. For spinors the appropriate
projectors are $\lambda^{\pm}=\frac{1}{2}\gamma^{\pm}\gamma^{\mp}\lambda$, so that
$\lambda=\lambda^{+}+\lambda^{-}$. At this step it is also convenient to fix
gauge condition. We employ light-cone gauge defined by setting ${\cal A}^{a\;+}=0$.
Then taking $\xi=\xi^{+}$ (so that $\delta {\cal A}^{\;+}=0$) the restricted light-cone
supersymmetry transformations (\ref{eq:2}) become
($\gamma_{\mu}^{\bot} = \gamma_{\mu} - n_{\mu}\gamma_{-}-n_{\mu}^{*}\gamma_{+}$,
where $n_{\mu}$ and $n_{\mu}^{*}$ project ``plus" and ``minus" components respectively)~\cite{LightConeSUSY}
\begin{eqnarray}
\delta^Q {\cal A}^{a\;\bot}_\mu&=&i\bar\xi^{+}\gamma^{\bot}_\mu\lambda^{a\;-},\qquad
\delta^Q A^a=\bar\xi^{+}\lambda^{a\;-},\qquad
\delta^Q B^a=\bar\xi^{+}\gamma_5\lambda^{a\;-},\nonumber\\
\delta^Q \lambda^{a\;-}&=&-\gamma^{-}\gamma^{\bot}_\mu\partial^{+}{\cal A}^{a\;\bot}_\mu
\xi^{+}
+i\gamma^{-}\partial^{+}(A^a+\gamma_5 B^a)\xi^{+},\label{SUSYTrLightCon}\\
\delta^Q\bar\lambda^{a\;-}&=&\bar\xi^{+}\gamma^{-}\gamma^{\bot}_\mu\partial^{+}
{\cal A}^{a\;\bot}_\mu+i\bar\xi^{+}\gamma^{-}\partial^{+}(A^a-\gamma_5 B^a).\nonumber
\end{eqnarray}
In light-cone gauge the restricted supersymmetry transformations form the off-shell
supersymmetry algebra. They are linear and form a closed algebra on the projected $+$-components
of fields, defined as components having maximal spin.

\subsection{Conformal twist-2 operators in $N=4$ SUSY YM}

Now, let us  introduce the local singlet (with respect to internal SU(4)-symmetry group)
conformal Wilson twist-2 operators appearing in this models for unpolarized and polarized
cases~\cite{CPO,BFKL2,SUSYCWI,OV}
\begin{eqnarray}
{\mathcal O}^{G}_{j,l}&=&
G_{+\mu}^{a\bot} (i\partial_+)^{l-1}
C^{5/2}_{j-1}\left(\frac{{\mathcal D}_+}{\partial_+}\right)
g^\bot_{\mu\nu}G_{\nu+}^{a\bot}\,,\label{ggn}\\
{\widetilde {\mathcal O}}^{G}_{j,l}&=&
G_{+\mu}^{a\bot} (i\partial_+)^{l-1}
C^{5/2}_{j-1}\left(\frac{{\mathcal D}_+}{\partial_+}\right)
\epsilon^\bot_{\mu\nu}G_{\nu+}^{a\bot}\,,\label{ggp}\\
{\mathcal O}^{\lambda}_{j,l}&=&
\frac12\bar \lambda^a_{+\,i} (i\partial_+)^l\gamma_+
C^{3/2}_j\left(\frac{{\mathcal D}_+}{\partial_+}\right)\lambda^{a\,i}_+\,,\label{qqn}\\
{\widetilde {\mathcal O}}^{\lambda}_{j,l}&=&
\frac12\bar \lambda^a_{+\,i} (i\partial_+)^l\gamma_+\gamma_5
C^{3/2}_j\left(\frac{{\mathcal D}_+}{\partial_+}\right)\lambda^{a\,i}_+\,,\label{qqp}\\
{\mathcal O}^{\phi}_{j,l}&=&
\bar\phi^a_r(i\partial_+)^{l+1}C^{1/2}_{j+1}\left(\frac{{\mathcal D}_+}{\partial_+}\right)\phi^a_r
\,,\label{ssn}
\end{eqnarray}
where $\phi =A+ i  B$ is a complex scalar field,
${\mathcal D}=\overrightarrow{\partial}- \overleftarrow{\partial}\!$,
$\partial=\overrightarrow{\partial}+\overleftarrow{\partial}\!$,
$g^{\perp}_{\mu\nu} = g_{\mu\nu}-n_{\mu}n^*_{\nu}-n_{\nu}n^*_{\mu}$,
$\epsilon^\perp_{\mu\nu} \equiv \epsilon^{\alpha\beta\rho\sigma}
g^\perp_{\alpha\mu} g^\perp_{\beta\nu} n^\ast_\rho n_\sigma$ and
$C^{\nu}_n(z)$ are Gegenbauer polynomials
\begin{equation}
C^{\nu}_n(z)=\frac{(-1)^n 2^n}{n!}\frac{\Gamma(n+\nu)}{\Gamma(\nu)}
\frac{\Gamma(n+2\nu)}{\Gamma(2n+2\nu)}(1-z^2)^{-\nu+1/2}
\frac{d^n}{dz^n}\left[(1-z^2)^{n+\nu-1/2}\right].
\end{equation}
In what follows we will restrict ourselves to the analysis of singlet
conformal operators, just introduced, and tensor operator to be introduced
in next section. However, in general, $N=4$ SYM has much richer content of twist-2
conformal operators, sitting in different representations of $SU(4)$-group. Besides bosonic
operators in other irreducible representations of $SU(4)$-group, we
can write down fermionic (by quantum numbers) operators formed by scalar-gluino and
gluon-gluino fields. While these fermionic operators were already present
in theories with less supersymmetry like $N=1$ SYM~\cite{BFKL2,SUSYCWI} and
Wess-Zumino model~\cite{OV}, here we encounter new type of operator - vector operator formed by scalar and gluon
fields. Under restricted supersymmetry transformations the conformal twist-2 operators
form a closed operator supermultiplet. A general procedure of constructing supermultiplets
of conformal operators is known for a long time \cite{BFKL2}. Recently, full set of twist-2 conformal
operators together with their transformations under restricted susy transformations was derived
for the case of $N=4$ SYM  in Ref.~\cite{BDKM}
and we refer interested reader to that paper.

As was already mentioned in introduction, we will be interested in the renormalization
properties of these operators. It should be noted, that in the singlet case there is mixing
between bosonic operators formed by gluon, gluino and scalar fields~(\ref{ggn})-(\ref{ssn}).
Also, in the non-forward kinematics, in contrast to forward case,
the operators (\ref{ggn})-(\ref{ssn}) will mix under renormalization
not only with each other, but also with the total derivatives of themselves.

\section{Supersymmetric Ward Identity in $N=4$ SUSY YM}

To begin with, let us introduce the following multiplicatively renormalized combinations of conformal operators
for unpolarized (Eqs.~(\ref{ggn}), (\ref{qqn}) and (\ref{ssn})) and polarized (Eqs.~(\ref{ggp}) and (\ref{qqp})) cases
\begin{eqnarray}
{\mathcal S}^1_{j,l}
&=&
6{\mathcal O}_{j,l}^{g}
+\frac{j}{2}{\mathcal O}_{j,l}^{\lambda}
+\frac{j(j+1)}{4}{\mathcal O}_{j,l}^{\phi}\,,
\label{scal1N4}\\
{\mathcal P}^1_{j,l}
&=&
6{\widetilde{\mathcal O}_{j,l}^{g}}
+\frac{j}{2}{\widetilde{\mathcal O}}_{j,l}^{\lambda}\,,
\label{pscal1N4}\\
{\mathcal S}^2_{j,l}
&=&
6{\mathcal O}_{j,l}^{g}
-\frac{1}{4}{\mathcal O}_{j,l}^{\lambda}
-\frac{(j+1)(j+2)}{12}{\mathcal O}_{j,l}^{\phi}\,,
\label{scal2N4}\\
{\mathcal P}^2_{j,l}
&=&
6{\widetilde{\mathcal O}}_{j,l}^{g}
-\frac{j+3}{2}{\widetilde{\mathcal O}}_{j,l}^{\lambda}\,,
\label{pscal2N4}\\
{\mathcal S}^3_{j,l}
&=&
6{\mathcal O}_{j,l}^{g}
-\frac{j+3}{2}{\mathcal O}_{j,l}^{\lambda}
+\frac{(j+2)(j+3)}{4}{\mathcal O}_{j,l}^{\phi}\,,
\label{scal3N4}
\end{eqnarray}
where the coefficients in front of operators can be found in a way, similar to~\cite{BLP,BFKL2}.
Here, we would like to note, that these linear combinations of conformal operators,
which are also the components of $N=4$ operator supermultiplet, renormalize multiplicatively
only in LO of perturbation theory. Beyond LO in dimensional reduction,
which preserves supersymmetry to rather high order in perturbation theory, used in this paper
they get additional rotations due to the
breakdown of superconformal symmetry. So, in general, it is only the constrains on anomalous
dimensions of our conformal operators  following from supersymmetric Ward identities, that
remain valid to all orders of perturbation theory.

To derive the supersymmetric Ward identities, relating anomalous dimensions of
singlet conformal operators to the anomalous dimension of some supersymmetry primary
operator, which renormalized multiplicatively, we need to know the action of
four restricted supersymmetry transformations on these singlet operators. Applying
four restricted supersymmetry transformations to our initial singlet operators we get
\footnote{Supersymmetric transformations for all operators in $N=4$ SYM can be found
in Appendix of Ref.\cite{BDKM}}:
($\delta^Q=\delta_4^Q\delta_3^Q\delta_2^Q\delta_1^Q$)
\begin{eqnarray}
\delta^Q {\mathcal S}^1_{j,l}&=&
\frac12(1-(-1)^j)
\bar\xi_4\gamma^\bot_\nu\gamma^-\xi_3
\bar\xi_2\gamma^\bot_\mu\gamma^-\xi_1
\ {\mathcal W}_{j-2,l}^{\mu\nu}\,,\label{COs1N4}\\
\delta^Q {\mathcal S}^2_{j,l}&=&
\frac12(1-(-1)^j)
\bar\xi_4\gamma^\bot_\nu\gamma^-\xi_3
\bar\xi_2\gamma^\bot_\mu\gamma^-\xi_1
\ {\mathcal W}_{j,l}^{\mu\nu}\,,\label{COs2N4}\\
\delta^Q {\mathcal S}^3_{j,l}&=&
\frac12(1-(-1)^j)
\bar\xi_4\gamma^\bot_\nu\gamma^-\xi_3
\bar\xi_2\gamma^\bot_\mu\gamma^-\xi_1
\ {\mathcal W}_{j+2,l}^{\mu\nu}\,,\label{COs3N4}\\
\delta^Q {\mathcal P}^1_{j,l}&=&
\frac12(1+(-1)^j)
\bar\xi_4\gamma^\bot_\nu\gamma^-\xi_3
\bar\xi_2\gamma^\bot_\mu\gamma^-\xi_1
\ \widetilde{\mathcal W}_{j-1,l}^{\mu\nu}\,,\label{COp1N4}\\
\delta^Q {\mathcal P}^2_{j,l}&=&
\frac12(1+(-1)^j)
\bar\xi_4\gamma^\bot_\nu\gamma^-\xi_3
\bar\xi_2\gamma^\bot_\mu\gamma^-\xi_1
\ \widetilde{\mathcal W}_{j+1,l}^{\mu\nu}\,,\label{COp2N4}
\end{eqnarray}
where
\begin{eqnarray}
{\mathcal W}^{\mu\nu}_{j,l} &=& \tau_{\perp}^{\mu\nu,\rho\sigma}{\rm tr}
\left\{
G^{+\perp}_{\rho} (i\partial_+)^{l-1}
C^{5/2}_{j-1}\left(\frac{{\mathcal D}_+}{\partial_+}\right)
G^{\perp+}_{\sigma}\,
\right\}, \\
\widetilde{\mathcal W}^{\mu\nu}_{j,l} &=& i \tau_{\perp}^{\mu\nu,\rho\sigma}{\rm tr}
\left\{
\widetilde{G}^{+\perp}_{\rho} (i\partial_+)^{l-1}
C^{5/2}_{j-1}\left(\frac{{\mathcal D}_+}{\partial_+}\right)
G^{\perp+}_{\sigma}\,
\right\},
\end{eqnarray}
$\tau_{\perp}^{\mu\nu ,\rho\sigma} = \frac{1}{2}\left(
g_{\perp}^{\mu\rho}g_{\perp}^{\nu\sigma} + g_{\perp}^{\mu\sigma}g_{\perp}^{\nu\rho}
-g_{\perp}^{\mu\nu}g_{\perp}^{\rho\sigma} \right)$.
Note, that here we kept only terms containing at the end operator we are interested in.

We see one general property, which is inherent to this set of transformation:
for bosonic operators with a certain parity (${\cal S}^i$ or ${\cal P}^i$) the index $j$ of
final operator ${\mathcal W}_{j,l}^{\mu\nu}$ changes on two units after each step.

Using restricted supersymmetry transformation above,
it is easy to find~\cite{SUSYCWI}, that the renormalized supersymmetric Ward identity in the regularization
scheme, preserving supersymmetry, has the following form
($\mbox{\boldmath$\cal S$}_{j,l}$ denotes vector of operators $S^1_{j,l}$, $S^2_{j,l}$
and $S^3_{j,l}$)
\begin{equation}
\langle [\mbox{\boldmath$\cal S$}_{jl}] \delta^Q {\cal X} \rangle
= - \langle \delta^Q [\mbox{\boldmath$\cal S$}_{jl}] {\cal X} \rangle
- \langle i [\mbox{\boldmath$\cal S$}_{jl}] (\delta^Q S) {\cal X} \rangle
\qquad\mbox{and}\quad
\langle \delta^Q [\mbox{\boldmath$\cal S$}_{jl}] {\cal X} \rangle = \mbox{finite}\,,\label{SCWI}
\end{equation}
where we used the fact, that renormalized action in supersymmetric regularization is invariant
with respect to supersymmetry transformations $\langle i [\mbox{\boldmath$\cal S$}_{jl}] (\delta^Q S) {\cal X} \rangle=0$.
As we already noted, beyond LO our operators (\ref{scal1N4}), (\ref{scal2N4}) and (\ref{scal3N4}) mix under
renormalization, so that the
renormalized operators are defined as (square brackets correspond to renormalized quantities)
\begin{equation}
\left[
\begin{array}{c}
{\cal S}^1\\ {\cal S}^2\\ {\cal S}^3
\end{array}
\right]_{jl}
= \sum_{k = 0}^{j}
\left(\begin{array}{ccc}
^{11}\!Z_{\cal S}& ^{12}\!Z_{\cal S}& ^{13}\!Z_{\cal S}\\
^{21}\!Z_{\cal S}& ^{22}\!Z_{\cal S}& ^{23}\!Z_{\cal S}\\
^{31}\!Z_{\cal S}& ^{32}\!Z_{\cal S}& ^{33}\!Z_{\cal S}
\end{array}\right)_{jk}
\left(\begin{array}{ccc}
Z_{\phi}^{-1} & 0  & 0 \\
0  & Z_{\phi}^{-1} & 0\\
0 & 0  & Z_{\phi}^{-1}
\end{array}\right)
\left(
\begin{array}{c}
{\cal S}^1\\ {\cal S}^2\\ {\cal S}^3
\end{array}
\right)_{kl}
\end{equation}
and as a consequence the renormalization group equation for these operators is given by
\begin{equation}
\frac{d}{d \ln \mu}
\left[
\begin{array}{c}
{\cal S}^1\\ {\cal S}^2\\ {\cal S}^3
\end{array}
\right]_{jl}=
 \sum_{k = 0}^{j}
\left(\begin{array}{ccc}
^{11}\!\gamma^{\cal S}& ^{12}\!\gamma^{\cal S}& ^{13}\!\gamma^{\cal S}\\
^{21}\!\gamma^{\cal S}& ^{22}\!\gamma^{\cal S}& ^{23}\!\gamma^{\cal S}\\
^{31}\!\gamma^{\cal S}& ^{32}\!\gamma^{\cal S}& ^{33}\!\gamma^{\cal S}\\
\end{array}\right)_{jk}
\left[
\begin{array}{c}
{\cal S}^1\\ {\cal S}^2\\ {\cal S}^3
\end{array}
\right]_{kl}\,.\label{MDGLAPN4}
\end{equation}
Now, from supersymmetric Ward identity (\ref{SCWI}) we get
($\sigma_k=\frac12(1-(-1)^k)$ and $Z_{jk} = 0$ for $k > j$)
\begin{equation}\label{RelADN4}
\sum_{k = 0}^{j} \sum_{k' = 0}^{k}
\left(
\begin{array}{lll}
 {^{11}\!Z}_{\cal S}  &  {^{12}\!Z}_{\cal S}&  {^{12}\!Z}_{\cal S} \\
 {^{21}\!Z}_{\cal S}  &  {^{22}\!Z}_{\cal S}&  {^{23}\!Z}_{\cal S} \\
 {^{31}\!Z}_{\cal S}  &  {^{32}\!Z}_{\cal S}&  {^{33}\!Z}_{\cal S} \\
\end{array}
\right)_{jk}
\sigma_k
\left(
\begin{array}{l}
\{ Z_{\cal W}^{-1} \}_{k - 2, k'}  \\
\{ Z_{\cal W}^{-1} \}_{k, k'}  \\
\{ Z_{\cal W}^{-1} \}_{k + 2, k'}
\end{array}
\right)
[{\cal W}_{k' l}]
= \mbox{finite}\,.
\end{equation}
$1/\epsilon$ poles in (\ref{RelADN4}) cancel, provided
\begin{eqnarray}
&&
\hspace*{-7mm}\!\sum_{k = 0}^{j} \left\{ {^{11}\!Z^{[1]}_{\cal S}} \right\}_{jk}
\!\!\sigma_k [{\cal W}_{k - 2, l}]
+\!\sum_{k = 0}^{j} \left\{ {^{12}\!Z^{[1]}_{\cal S}} \right\}_{jk}
\!\!\sigma_k [{\cal W}_{k, l}]
+ \!\sum_{k = 0}^{j} \left\{ {^{13}\!Z^{[1]}_{\cal S}} \right\}_{jk}
\!\!\sigma_k [{\cal W}_{k + 2,l}]
= \sigma_j \!\sum_{k = 0}^{j} \left\{ {Z^{[1]}_{\cal W}} \right\}_{j - 2, k}
[{\cal W} _{kl}] ,\nonumber\\
&&
\hspace*{-7mm}\!\sum_{k = 0}^{j} \left\{ {^{21}\!Z^{[1]}_{\cal S}} \right\}_{jk}
\!\!\sigma_k [{\cal W} _{k - 2, l}]
+\!\sum_{k = 0}^{j} \left\{ {^{22}\!Z^{[1]}_{\cal S}} \right\}_{jk}
\!\!\sigma_k [{\cal W} _{k, l}]
+ \!\sum_{k = 0}^{j} \left\{ {^{23}\!Z^{[1]}_{\cal S}} \right\}_{jk}
\!\!\sigma_k [{\cal W} _{k + 2, l}]
= \sigma_j \!\sum_{k = 0}^{j} \left\{ {Z^{[1]}_{\cal W}} \right\}_{j,k}
[{\cal W} _{kl}] ,\nonumber\\
&&
\hspace*{-7mm}\!\sum_{k = 0}^{j} \left\{ {^{31}\!Z^{[1]}_{\cal S}} \right\}_{jk}
\!\!\sigma_k [{\cal W} _{k - 2, l}]
+\!\sum_{k = 0}^{j} \left\{ {^{32}\!Z^{[1]}_{\cal S}} \right\}_{jk}
\!\!\sigma_k [{\cal W} _{k, l}]
+ \!\sum_{k = 0}^{j} \left\{ {^{33}\!Z^{[1]}_{\cal S}} \right\}_{jk}
\!\!\sigma_k [{\cal W} _{k + 2, l}]
= \sigma_j \!\sum_{k = 0}^{j} \left\{ {Z^{[1]}_{\cal W}} \right\}_{j + 2, k}
[{\cal W} _{kl}] .\nonumber
\end{eqnarray}
Taking into account linear independence of  operators $[{\cal W} _{kl}]$ we finally get
the following relations on anomalous dimensions of conformal operators
\begin{eqnarray}\label{S1N4}
&&
{}^{11}\!\gamma^{\cal S}_{2n + 5, 2m + 1}+{}^{12}\!\gamma^{\cal S}_{2n + 5, 2m - 1}
+{}^{13}\!\gamma^{\cal S}_{2n + 5, 2m - 3}= \gamma^{\cal W}_{2n + 3, 2m - 1} ,
\quad\quad m\leq n+2\,, \\ \label{S2N4}
&&
{^{21}\!\gamma}^{\cal S}_{2n + 3, 2m + 1}+{^{22}\!\gamma}^{\cal S}_{2n + 3, 2m - 1}
+{^{23}\!\gamma}^{\cal S}_{2n + 3, 2m - 3}= \gamma^{\cal W}_{2n + 3, 2m - 1} ,
\quad\quad m\leq n+1\,, \\\label{S3N4}
&&
{^{31}\!\gamma}^{\cal S}_{2n + 1, 2m + 1}+{^{32}\!\gamma}^{\cal S}_{2n + 1, 2m - 1}
+{^{33}\!\gamma}^{\cal S}_{2n + 1, 2m - 3}= \gamma^{\cal W}_{2n + 3, 2m - 1} ,
\quad\quad m\leq n
\end{eqnarray}
and
\begin{eqnarray}
\label{DokN4}
&&
{^{12}\!\gamma}^{\cal S}_{2n + 1, 2n + 1}
+{^{13}\!\gamma}^{\cal S}_{2n + 1, 2n - 1}=0\,,\quad\quad\qquad\quad\;\,
{^{13}\!\gamma}^{\cal S}_{2n + 1, 2n + 1}=0\,,\\
&&
{^{22}\!\gamma}^{\cal S}_{2n + 1, 2n + 1}
+{^{23}\!\gamma}^{\cal S}_{2n + 1, 2n - 1}
= \gamma^{\cal W}_{2n + 1, 2n + 1}\,,\quad\quad
{^{23}\!\gamma}^{\cal S}_{2n + 1, 2n + 1}=0\,,\\
&&
{^{32}\!\gamma}^{\cal S}_{2n + 1, 2n + 1}
+{^{33}\!\gamma}^{\cal S}_{2n + 1, 2n - 1}
= \gamma^{\cal W}_{2n + 3, 2n + 1}\,,\quad\quad
{^{33}\!\gamma}^{\cal S}_{2n + 1, 2n + 1}
= \gamma^{\cal W}_{2n + 3, 2n + 3}\,.
\end{eqnarray}

Now, let us turn to polarized case. All the steps one need to perform here are
in one to one correspondence with unpolarized case. First we write down the
supersymmetric Ward identity
($\mbox{\boldmath$\cal P$}_{j,l}$ denotes vector of operators $P^1_{j,l}$~(\ref{pscal1N4})
and $P^2_{j,l}$~(\ref{pscal2N4}))
\begin{equation}
\langle [\mbox{\boldmath$\cal P$}_{jl}] \delta^Q {\cal X} \rangle
= - \langle \delta^Q [\mbox{\boldmath$\cal P$}_{jl}] {\cal X} \rangle
- \langle i [\mbox{\boldmath$\cal P$}_{jl}] (\delta^Q P) {\cal X} \rangle
\qquad\mbox{and}\quad
\langle \delta^Q [\mbox{\boldmath$\cal P$}_{jl}] {\cal X} \rangle = \mbox{finite}\,.\label{SCWIp}
\end{equation}
The operators (\ref{pscal1N4}) and (\ref{pscal2N4}) mix under renormalization and
thus we define renormalized operators as (square brackets correspond to renormalized quantities)
\begin{equation}
\left[
\begin{array}{c}
{\cal P}^1\\ {\cal P}^2
\end{array}
\right]_{jl}
= \sum_{k = 0}^{j}
\left(\begin{array}{ccc}
^{11}\!Z_{\cal P}& ^{12}\!Z_{\cal P}\\
^{21}\!Z_{\cal P}& ^{22}\!Z_{\cal P}
\end{array}\right)_{jk}
\left(\begin{array}{ccc}
Z_{\phi}^{-1} & 0 \\
0  & Z_{\phi}^{-1}
\end{array}\right)
\left(
\begin{array}{c}
{\cal P}^1\\ {\cal P}^2
\end{array}
\right)_{kl}.
\end{equation}
The renormalization group equation for these operators is given by
\begin{equation}
\frac{d}{d \ln \mu}
\left[
\begin{array}{c}
{\cal P}^1\\ {\cal P}^2
\end{array}
\right]_{jl}=
 \sum_{k = 0}^{j}
\left(\begin{array}{ccc}
^{11}\!\gamma^{\cal P}& ^{12}\!\gamma^{\cal P}\\
^{21}\!\gamma^{\cal P}& ^{22}\!\gamma^{\cal P}
\end{array}\right)_{jk}
\left[
\begin{array}{c}
{\cal P}^1\\ {\cal P}^2
\end{array}
\right]_{kl}\,.\label{MDGLAPN4p}
\end{equation}
From supersymmetric Ward identity (\ref{SCWIp}) we get
($\bar\sigma_k=\frac12(1+(-1)^k)$ and $Z_{jk} = 0$ for $k > j$)
\begin{equation}\label{RelADN4p}
\sum_{k = 0}^{j} \sum_{k' = 0}^{k}
\left(
\begin{array}{lll}
 {^{11}\!Z}_{\cal P}  &  {^{12}\!Z}_{\cal P}\\
 {^{21}\!Z}_{\cal P}  &  {^{22}\!Z}_{\cal P}\\
\end{array}
\right)_{jk}
\bar\sigma_k
\left(
\begin{array}{l}
\{ Z_{\widetilde{\cal W}}^{-1} \}_{k - 1, k'}  \\
\{ Z_{\widetilde{\cal W}}^{-1} \}_{k + 1, k'}
\end{array}
\right)
[{\widetilde{\cal W}}_{k' l}]
= \mbox{finite}\,.
\end{equation}
$1/\epsilon$ poles in (\ref{RelADN4p}) cancel, provided
\begin{eqnarray}
&&
\hspace*{-7mm}\sum_{k = 0}^{j} \left\{ {^{11}\!Z^{[1]}_{\cal P}} \right\}_{jk}
\!\!\bar\sigma_k [{\widetilde{\cal W}}_{k - 1, l}]
+\sum_{k = 0}^{j} \left\{ {^{12}\!Z^{[1]}_{\cal P}} \right\}_{jk}
\!\!\bar\sigma_k [{\widetilde{\cal W}}_{k + 1, l}]
= \bar\sigma_j \sum_{k = 0}^{j} \left\{ {Z^{[1]}_{\widetilde{\cal W}}} \right\}_{j - 1, k}
[{\widetilde{\cal W}} _{kl}] ,\nonumber\\
&&
\hspace*{-7mm}\sum_{k = 0}^{j} \left\{ {^{21}\!Z^{[1]}_{\cal P}} \right\}_{jk}
\!\!\bar\sigma_k [{\widetilde{\cal W}} _{k - 1, l}]
+\sum_{k = 0}^{j} \left\{ {^{22}\!Z^{[1]}_{\cal P}} \right\}_{jk}
\!\!\bar\sigma_k [{\widetilde{\cal W}} _{k + 1, l}]
= \bar\sigma_j \sum_{k = 0}^{j} \left\{ {Z^{[1]}_{\widetilde{\cal W}}} \right\}_{j + 1,k}
[{\widetilde{\cal W}} _{kl}] .\nonumber
\end{eqnarray}
Taking into account linear independence of operators
$[{\widetilde{\cal W}} _{kl}]$ we finally get the following
relations
\begin{eqnarray}\label{P1N4}
&&
{}^{11}\!\gamma^{\cal P}_{2n + 2, 2m}+{}^{12}\!\gamma^{\cal P}_{2n + 2, 2m - 2}
= \gamma^{\widetilde{\cal W}}_{2n, 2m - 2} \,,  \qquad\qquad m\leq n+1\,,\\
\label{P2N4}
&&
{^{21}\!\gamma}^{\cal P}_{2n, 2m}+{^{22}\!\gamma}^{\cal P}_{2n, 2m - 2}
= \gamma^{\widetilde{\cal W}}_{2n, 2m - 2}\,,\qquad\qquad\qquad m\leq n
\end{eqnarray}
and
\begin{eqnarray}
\label{DokN4p}
&&
{^{12}\!\gamma}^{\cal P}_{2n, 2n}=0\,,\qquad\qquad\qquad\quad\;\,
{^{22}\!\gamma}^{\cal P}_{2n, 2n}
= \gamma^{\widetilde{\cal W}}_{2n, 2n}\,.
\end{eqnarray}
Note, that both operators $W_{j,k}$ and ${\widetilde W}_{j,k}$ are components of the
same operator supermultiplet and thus have the same single anomalous dimension
(with shifted argument).

To obtain from these results  the corresponding relations for anomalous dimensions
of conformal operators considered earlier, one just need to use the following transition formulae
among anomalous dimensions of conformal operators (\ref{scal1N4})-(\ref{scal3N4})
and anomalous dimensions of operators (\ref{ggn})-(\ref{ssn}):
\begin{equation}
\left(\begin{array}{ccc}
^{11}\!\gamma^{\cal P}& ^{12}\!\gamma^{\cal P}\\[2mm]
^{21}\!\gamma^{\cal P}& ^{22}\!\gamma^{\cal P}
\end{array}\right)
=
\left(\begin{array}{ccc}
\scriptstyle 6 & \frac{j}{2}    \\[2mm]
\scriptstyle 6 & {\scriptstyle-}\frac{j+3}{2}
\end{array}\right)
\left(\begin{array}{ccc}
^{gg}\tilde\gamma       & ^{g\lambda}\tilde\gamma   \\[2mm]
^{\lambda g}\tilde\gamma& ^{\lambda\lambda}\tilde\gamma
\end{array}\right)
\left(\begin{array}{ccc}
\scriptstyle 6 & \frac{k}{2}   \\[2mm]
\scriptstyle 6 & {\scriptstyle-}\frac{k+3}{2}
\end{array}\right)^{\!\!-1}\,,\label{NumToF}
\end{equation}

\begin{equation}
\left(\begin{array}{ccc}
^{11}\!\gamma^{\cal S}\! & \! ^{12}\!\gamma^{\cal S}\! & \! ^{13}\!\gamma^{\cal S}\\[3mm]
^{21}\!\gamma^{\cal S}\! & \! ^{22}\!\gamma^{\cal S}\! & \! ^{23}\!\gamma^{\cal S}\\[3mm]
^{31}\!\gamma^{\cal S}\! & \! ^{32}\!\gamma^{\cal S}\! & \! ^{33}\!\gamma^{\cal S}
\end{array}\right)
=
\left(\begin{array}{ccc}
\scriptstyle 6 \! & \! \frac{j}{2}    \! & \! \frac{j(j+1)}{4}\\[2mm]
\scriptstyle 6 \! & \! {\scriptstyle-}\frac{1}{4}   \! & \! {\scriptstyle-}\frac{(j+1)(j+2)}{12}\\[2mm]
\scriptstyle 6 \! & \! {\scriptstyle-}\frac{j+3}{2} \! & \! \frac{(j+2)(j+3)}{4}
\end{array}\right)
\left(\begin{array}{ccc}
^{gg}\gamma       \! & \! ^{g\lambda}\gamma      \! & \! ^{g\phi}\gamma\\[3mm]
^{\lambda g}\gamma\! & \! ^{\lambda\lambda}\gamma\! & \! ^{\lambda\phi}\gamma\\[3mm]
^{\phi g}\gamma   \! & \! ^{\phi\lambda}\gamma   \! & \! ^{\phi\phi}\gamma
\end{array}\right)
\left(\begin{array}{ccc}
\scriptstyle 6 \! & \! \frac{k}{2}    \! & \! \frac{k(k+1)}{4}\\[2mm]
\scriptstyle 6 \! & \! {\scriptstyle-}\frac{1}{4}   \! & \! {\scriptstyle-}\frac{(k+1)(k+2)}{12}\\[2mm]
\scriptstyle 6 \! & \! {\scriptstyle-}\frac{k+3}{2} \! & \! \frac{(k+2)(k+3)}{4}
\end{array}\right)^{\!\!-1}\!\!\!\!.
\end{equation}
These relations allow us to operate both with anomalous dimensions of conformal operators
(\ref{ggn})-(\ref{ssn})
or, equivalently, with anomalous dimensions of multiplicatively renormalized operators
(\ref{scal1N4})-(\ref{scal3N4}).

\section{Universal non-forward AD in $N=4$ SUSY YM.}

Now, let us proceed with the determination of universal
non-forward anomalous dimension ${\gamma}^{uni} _{j,k}=
\gamma^{\cal W}_{j,k}$. The simplest way to do it is to use the
formalism of broken conformal Ward identities (CWI)
Ref.~\cite{MPhi3,MQCDNS}. The basic idea of this method lies in
the relation between scale and special conformal anomalies of
conformal operators. It was found that non-diagonal part
$\gamma^{\rm ND}$ of the complete anomalous dimensions matrix
\begin{equation}
\gamma_{j,k} = \gamma^{\rm D}_{j}\delta_{j,k} + \gamma^{\rm ND}_{j,k}, \qquad\qquad \
\gamma=\frac{\alpha_s N_c}{2 \pi}\gamma^{(0)}+
\left(\frac{\alpha_s N_c}{2 \pi}\right)^2\gamma^{(1)}+...
\end{equation}
arises entirely due to the violation of special conformal symmetry. Moreover,
within this framework we have a relation between scale anomalous
dimension matrix in $n$-th order of perturbation theory and
matrices of scale and special conformal anomalies in
$(n\!-\!1)$-order of perturbation theory. As a consequence the
calculation of leading nondiagonal part of nonforward anomalous
dimensions matrix, which is nonzero starting from two-loop order,
could be reduced to calculation of more simple one-loop diagrams.

Up to the moment there are results for nondiagonal parts of nonforward anomalous dimension
matrices of conformal operators calculated in QED~\cite{BMQED}, QCD~\cite{BMQCD},
$N=1$ supersymmetric Yang-Mills theory~\cite{SUSYCWI} and in supersymmetric
Wess-Zumino model~\cite{OV}, both for polarized and unpolarized cases.
The results in  $N=1$ SYM could be easily derived from already known QCD result via simple
identification of colour Casimir operators: $C_A = C_F = 2N_f T_F = N_c$.  In $N=4$ SYM the situation
is slightly more complicated. Besides correspondingly adjusted map of colour Casimir operators:
$C_A = C_F = \frac{1}{2}N_f T_F = N_c$, where we accounted for four Majorana fermions compared
to one in $N=1$ SYM, we need to consider scalar fields $\phi$, which are absent in $N=1$ SYM.
The presence of scalar fields results in introduction of one additional conformal
operator~(\ref{scal3N4}) and
increases the size of anomalous dimension matrix in unpolarized case. So, in general, in unpolarized case
we need to compute 5 new conformal anomalies.  However, as we already noted, to find leading nondiagonal
part of nonforward anomalous dimensions matrix all we need to know is one-loop matrix of special
conformal anomalies (matrix of scale anomalies is diagonal at one-loop order and is known for a long time).
A simple diagram analysis shows, that the only nontrivial contribution to special conformal anomaly matrix,
not related to one-loop scale anomaly matrix and not already computed in QCD, resides in $^{\phi\phi}\gamma$
and $^{g\phi}\gamma$.

On the contrary, in polarized case we have only two conformal operators: (\ref{ggp})
and (\ref{qqp}), which are similar to polarized operators in $N=1$ SYM and QCD. In this case we do not
need to perform any new calculations and we can use polarized case to extract universal
non-diagonal part of non-forward anomalous dimension matrix from already known
results for non-forward polarized anomalous dimension matrix in QCD from Refs.~\cite{MQCDNS,BMQED,BMQCD}
and results for leading order forward polarized anomalous dimension matrix in
$N=4$ SYM from Refs.~\cite{LN4,KL}. As a result we obtain the following final expressions
for leading non-diagonal parts of non-forward polarized anomalous dimension matrix in $N=4$ SYM
$(j > k )$:
\begin{eqnarray}
{^{\lambda\lambda}{\widetilde\gamma}}_{jk}^{{\rm ND}(1)}
&=&
-\left(
{^{\lambda\lambda}{\widetilde\gamma}}_{j}^{(0)} - {^{\lambda\lambda}{\widetilde\gamma}}_{k}^{(0)}
\right)
\left(
d_{jk}
 {^{\lambda\lambda}{\widetilde\gamma}}_{k}^{(0)}
- {^{\lambda\lambda}g}_{jk}
\right) \\
&&\qquad\qquad\qquad\qquad\quad -
\left(
{^{\lambda g}{\widetilde\gamma}}_{j}^{(0)} - {^{\lambda g}{\widetilde\gamma}}_{k}^{(0)}
\right) d_{jk}
{^{g\lambda}{\widetilde\gamma}}_{k}^{(0)}
+ {^{\lambda g}{\widetilde\gamma}}_{j}^{(0)} {^{g\lambda}g}_{jk}\,, \nonumber\label{ndadll}\\
{^{\lambda g}{\widetilde\gamma}}_{jk}^{{\rm ND}(1)}
&=&
-\left(
{^{\lambda g}{\widetilde\gamma}}_{j}^{(0)} - {^{\lambda g}{\widetilde\gamma}}_{k}^{(0)}
\right)
d_{jk}
{^{gg}{\widetilde\gamma}}_{k}^{(0)}
- \left(
{^{\lambda\lambda}{\widetilde\gamma}}_{j}^{(0)} - {^{\lambda\lambda}{\widetilde\gamma}}_{k}^{(0)}
\right)
d_{jk} {^{\lambda g}{\widetilde\gamma}}_{k}^{(0)} \\
&&\qquad\qquad\qquad\qquad\quad +
{^{\lambda g}{\widetilde\gamma}}_{j}^{(0)} {^{gg}g}_{jk}
-
{^{\lambda\lambda}g}_{jk} {^{\lambda g}{\widetilde\gamma}}_{k}^{(0)} , \nonumber\label{ndadlg}\\
{^{g\lambda}{\widetilde\gamma}}_{jk}^{{\rm ND}(1)}
&=&
-\left(
{^{g\lambda}{\widetilde\gamma}}_{j}^{(0)} - {^{g\lambda}{\widetilde\gamma}}_{k}^{(0)}
\right) d_{jk}
{^{\lambda\lambda}{\widetilde\gamma}}_{k}^{(0)}
-
\left(
{^{gg}{\widetilde\gamma}}_{j}^{(0)} - {^{gg}{\widetilde\gamma}}_{k}^{(0)}
\right) d_{jk}
{^{g\lambda}{\widetilde\gamma}}_{k}^{(0)} \\
&&\qquad\qquad\qquad\qquad\quad +
{^{g\lambda}{\widetilde\gamma}}_{j}^{(0)} {^{\lambda\lambda}g}_{jk}
-
{^{gg}g}_{jk} {^{g\lambda}{\widetilde\gamma}}_{k}^{(0)}
+
\left(
{^{gg}{\widetilde\gamma}}_{j}^{(0)} - {^{\lambda\lambda}{\widetilde\gamma}}_{k}^{(0)}
\right)
{^{g\lambda}g}_{jk} \,, \nonumber\label{ndadgl}\\
{^{gg}{\widetilde\gamma}}_{jk}^{{\rm ND}(1)}
&=&
-\left(
{^{gg}{\widetilde\gamma}}_{j}^{(0)} - {^{gg}{\widetilde\gamma}}_{k}^{(0)}
\right)
\left(
d_{jk}
{^{gg}{\widetilde\gamma}}_{k}^{(0)}
- {^{gg}g}_{jk}
\right) \\
&&\qquad\qquad\qquad\qquad\quad -
\left(
{^{g\lambda}{\widetilde\gamma}}_{j}^{(0)} - {^{g\lambda}{\widetilde\gamma}}_{k}^{(0)}
\right) d_{jk}
{^{\lambda g}{\widetilde\gamma}}_{k}^{(0)}
-
{^{g\lambda}g}_{jk}{^{\lambda g}{\widetilde\gamma}}_{k}^{(0)}. \nonumber\label{ndadgg}
\end{eqnarray}
where $d_{jk} = b(j,k)/a(j,k)$, $g_{jk}= w_{jk}/a(j,k)$ and
\begin{eqnarray}
a(j,k) &=& 2(j-k)(j+k+3), \nonumber \\
b(j,k) &=&  -[1+(-1)^{j-k}](2k + 3),~~\mbox{if $j>k$ and even\,; 0 otherwise}\,. \nonumber
\end{eqnarray}
The quantities $w_{j,k}$ were computed in Refs.~\cite{MQCDNS,BMQED,BMQCD} and are given by
\begin{eqnarray}
^{gg}w_{j,k} &=& -2 [1+(-1)^{j-k}]\theta_{j-2,k}(3+2k)  \nonumber \\
&&\times
\left(
2A_{jk} +(A_{jk}-S_1(j+1))\left[\frac{(j)_4}{(k)_4}-1\right] + 2(j-k)(j+k+3)\frac{1}{(k)_4}
\right), \\
^{\lambda\lambda}w_{jk} &=& -2[1+(-1)^{j-k}]\theta_{j-2,k}(3+2k) \nonumber \\
&&\times
\left(
2A_{jk} + (A_{jk}-S_1(j+1))\frac{(j-k)(j+k+3)}{(k+1)(k+2)}
\right), \\
^{g\lambda}w_{jk} &=& -2[1+(-1)^{j-k}]\theta_{j-2,k}(3+2k)\frac{1}{6}\frac{(j-k)(j+k+3)}{(k+1)(k+2)}\,, \\
^{\lambda g}w_{jk} &=& 0\,,
\end{eqnarray}
where
$\theta_{j,k}$ is equal to $1$, if $j>k$ and $0$ otherwise
and
\begin{eqnarray}
A_{jk} &=& S_1\left(\frac{j+k+2}{2}\right) - S_1\left(\frac{j-k-2}{2}\right) + 2S_1(j-k-1) - S_1(j+1).
\end{eqnarray}
Here $S_k(n)$ denotes harmonic sum
\begin{equation}
S_i(j)=\sum^{j}_{m=1}\frac{1}{m^i}\;,\qquad
S_{-i}(j)=(-1)^j \sum^{j}_{m=1}\frac{(-1)^{m}}{m^i}\;,\qquad
S_{-2,1}(j)=(-1)^j \sum^{j}_{m=1}\frac{(-1)^{m}}{m^2}\,S_1(m)\label{HarmSum}
\end{equation}
and $(a)_k = \frac{\Gamma (a+k)}{\Gamma (a)}$
stands for Pochhammer symbol. The leading order forward anomalous dimensions of conformal operators, entering formula
above are given by \cite{LN4,KL}
\begin{eqnarray}
{^{\lambda\lambda}\!{\widetilde\gamma}}_{j}^{(0)}
&=&
4 \left(
\frac{1}{( j + 1 )( j + 2 )} + S_1(j+1)
\right),
\\
{^{\lambda g}\!{\widetilde\gamma}}_{j}^{(0)}
&=&
-\frac{48}{( j + 1 )( j + 2 )}\;,
\\
{^{g\lambda}\!{\widetilde\gamma}}_{j}^{(0)}
&=&
-\frac{j ( j + 3 )}{3( j + 1 )( j + 2 )}\;,
\\
{^{gg}\!{\widetilde\gamma}}_{j}^{(0)}
&=&
4 S_1(j+1) + \frac{8}{( j + 1 )( j + 2 )}\;.
\end{eqnarray}

Collecting all contributions together and using the relations~(\ref{NumToF}) and
(\ref{P1N4}) (or (\ref{P2N4})) we find the following expression for universal non-forward
anomalous dimension in $N=4$ SYM
\begin{eqnarray}
{}^{uni}\gamma^{\,{\rm ND}(1)}_{j,k}&=&\gamma^{\widetilde {\cal W}}_{j-2, k-2}\;=\;
4\frac{\bigl(1+(-1)^{j-k}\bigr)(2k+1)\bigl(S_1(j)-S_1(k)\bigr)}
{(k-1)_4(j-k)(j+k+1)}\times\nonumber\\[3mm]
&\hspace*{-17mm}\times&
\hspace*{-10mm}\biggl\{\left[S_1\!\!\left(\frac{j-k}{2}\right)-2S_1(j-k)-S_1\!\!\left(\frac{j+k}{2}\right)\right]
\Bigl((j-1)_4+(k-1)_4\Bigr)\nonumber\\
&& + 2(j-1)_4S_1(j)+2(k-1)_4S_1(k)\biggr\}.\label{N4Res}
\end{eqnarray}

Now, combining Eqs.~(\ref{P1N4}) and (\ref{P2N4}) together we get
\begin{equation}\label{P1P2T}
{}^{11}\!\gamma^{\cal P}_{2n + 2, 2m}+{}^{12}\!\gamma^{\cal P}_{2n + 2, 2m - 2}
- {^{21}\!\gamma}^{\cal P}_{2n, 2m}-{^{22}\!\gamma}^{\cal P}_{2n, 2m - 2}
=0\,.
\end{equation}
Taking $m=n-1$ it is easy to check using Eq.~(\ref{NumToF}), that
our nonforward anomalous dimensions satisfy this relation.
Moreover we can check relation between the NLO non-forward
anomalous dimension and NLO forward anomalous dimension from
Ref.~\cite{KLV} \footnote{In Ref.~\cite{KLV} the results of
calculation were written in DRED scheme, however the coupling
constant was kept in ${\mathrm{\overline{MS}}}$ scheme. To obtain
a result fully expressed in DRED quantities, one needs to perform
the following coupling constant redefinition for coupling constant
$\alpha_{\mathrm {DRED}}=\alpha_{\mathrm {\overline {MS}}}+
N_c/(12\pi)\left(\alpha_{\mathrm {\overline {MS}}}\right)^2$. It
means, that we need to add to NLO anomalous dimension from
Ref.~\cite{KLV} the corresponding LO expression from
Refs.~\cite{LN4,KL}, that is in present paper we are using
$\gamma^{(1)}_j=\left[\gamma^{(1)}_j\right]_{\mathrm{Ref.\cite{KLV}}}
-1/3\,\gamma^{(0)}_j$}.
Namely from Eq.~(\ref{P1P2T}) we have
\begin{eqnarray}
{}^{11}\!\gamma^{\cal P}_{2n + 2, 2n}+{}^{12}\!\gamma^{\cal P}_{2n + 2, 2n - 2}
- {^{21}\!\gamma}^{\cal P}_{2n, 2n}-{^{22}\!\gamma}^{\cal P}_{2n, 2n - 2}
=0\, ,&\qquad \mathrm{for}& m=n\, ,\label{P1P2D1}\\
{}^{11}\!\gamma^{\cal P}_{2n, 2n}+{}^{12}\!\gamma^{\cal P}_{2n, 2n - 2}
-{^{22}\!\gamma}^{\cal P}_{2n - 2, 2n - 2}
=0\, ,&\qquad \mathrm{for}& m=n+1\, .\label{P1P2D2}
\end{eqnarray}
Indeed, taking explicit expressions for NLO forward anomalous dimensions from Ref.~\cite{KLV}
(to adapt these results to normalization used in present paper one needs to
multiply the results of Ref.~\cite{KLV} by $(-\frac12)\,$,
shift the value
of momentum $j$ by unity $j=j+1$ and then make the following substitutions:
${}^{\lambda g}\tilde\gamma_j={}^{\lambda g}\tilde\gamma_j\, \frac{6}{j}$ and
${}^{g\lambda}\tilde\gamma_j={}^{g\lambda}\tilde\gamma_j\,\frac{j}{6}{}$)
\begin{eqnarray}
 {}^{11}\!\gamma^{\cal P}_{j, j}&=&
 {}^{22}\!\gamma^{\cal P}_{j-2, j-2}\ =\
 -4\bigl(S_3(j)+S_{-3}(j)\bigr)+8 S_1(j)\bigl(S_2(j)+S_{-2}(j)\bigr)- 8S_{-2,1}(j)\,,\qquad\\
 {}^{21}\!\gamma^{\cal P}_{j, j}&=&
\frac{24 S_1(j)}{j(j+1)(j+2)}-\frac{8(2j+3)}{(j+1)^2(j+2)^2}\,.
\end{eqnarray}
where $S_i(j)$ stands for harmonic sum defined in Eq.~(\ref{HarmSum}), one can
easily verify, that above relations hold true. Note, that
Eq.~(\ref{P1P2D1}) contains a contribution from forward
non-diagonal anomalous dimension ${}^{21}\gamma_{2n,2n}$. The
appearance of nondiagonal part of forward anomalous dimension
matrix~(\ref{NumToF}) is related to the breaking of superconformal symmetry, what is
explicitly demonstrated by Eq.~(\ref{P1P2D1}).

Finally, having in mind possible future tests of AdS/CFT correspondence, we would like to consider
one particular limit of our result Eq.~(\ref{N4Res}). As this result is  meaningful only
for $j>k$ we parameterize the limit of large $j$ and $k$ as follows:
\begin{equation}
j=n\,m\,, \qquad\qquad k=m\,.
\end{equation}
Now, using $S_1(j)=\Psi(j+1)-\Psi(1)\approx \ln(j)$ for large $j$, it is easy to find the
following asymptotic behaviour of Eq.~(\ref{N4Res}) (for even $j-k$)
\begin{eqnarray}
{}^{uni}\gamma^{{\rm ND}(1)}_{n\,m,m}=\frac{16}{m}\frac{\ln(n)}{n^2-1}
\left[n^4\ln\left(\frac{n^2}{n^2-1}\right)+\ln\left(\frac{1}{n^2-1}\right)\right].
\end{eqnarray}

\section{Conclusion}

In the present paper we found a closed analytical expression for NLO
universal non-forward anomalous dimension of Wilson twist-2 operators in
$N=4$ supersymmetric Yang-Mills theory.
To derive this result it was sufficient to know constrains on anomalous
dimensions of conformal operators, following from supersymmetric Ward
identities, together with already known results for anomalous dimensions
of Wilson twist-2 operators in QCD. Moreover, supersymmetric Ward identities
allowed us to find the relation of non-forward anomalous dimension
determined here with already known forward anomalous dimension of
Wilson twist-2 operators in  $N=4$ SYM theory. This relation may serve
as an additional check of correctness of the results obtained
both for forward and non-forward anomalous dimensions of conformal operators
in this model up to next-to-leading order in perturbation theory. Finally, we considered the limit
of our result for non-forward anomalous dimension, when momenta $j$ and $k$
turn to infinity, while their ratio is being fixed. In this limit non-forward
anomalous dimension scales as $1/j$, where $j$ is Lorentz spin.

\vspace{1cm}
{\large \bf Acknowledgments}

\vspace*{0.5cm}
\noindent
We are grateful to L.N.~Lipatov  for numerous discussions on the subject of this
paper. The work of A.O. was supported by the National Science Foundation under grant PHY-0244853 and by
the US Department of Energy under grant DE-FG02-96ER41005. The work of V.V.
is supported by grants INTAS 00-366 and RSGSS-1124.2003.2.

\end{document}